\documentclass[a4paper]{PoS}
\usepackage{graphicx,amssymb}
\usepackage[T1]{fontenc}
\usepackage{aecompl}
\usepackage{verbatim}
\usepackage{color}
\usepackage{multirow}
\usepackage{fixmath}
\usepackage{rotating}
\usepackage{pbox}
\usepackage{pdflscape}

\title{Linear and circular polarization in GRB afterglows}

\ShortTitle{Polarization in GRB afterglows}

\author{\speaker{Lara Nava}%
       \\
       The Hebrew University of Jerusalem, Jerusalem, 91904, Israel\\
       E-mail: \email{lara.nava@mail.huji.ac.il}}

\author{Ehud Nakar\\
        Tel Aviv University, Tel Aviv, 69978, Israel\\
        E-mail: \email{udini@wise.tau.ac.il}}

\author{Tsvi Piran\\
        The Hebrew University of Jerusalem, Jerusalem, 91904, Israel\\
        E-mail: \email{tsvi.piran@mail.huji.ac.il}}

\abstract{A certain degree of linear polarization has been measured in several GRB afterglows. More surprisingly, circular polarization has been recently measured in GRB121024A. For synchrotron emission, the polarization level depends on: (i) the local magnetic field orientation (ii) the geometry of the emitting region with respect to the line of sight and (iii) the electron pitch-angle distribution. For this reason, polarization measurements are a valuable tool to probe afterglow micro-physics. We present numerical estimates of linear and circular polarization for different configurations (i.e., magnetic fields, geometries and pitch-angle distributions). For each different scenario, we study the conditions for reaching the maximum and minimum linear and circular polarization and provide their values. We discuss the implication of our results to the micro-physics of GRB afterglows in view of recent polarization measurements.}

\FullConference{Swift: 10 Years of Discovery\\
		 2-5 December 2014\\
		 La Sapienza University, Rome, Italy}

\begin{document}

\section{Introduction}
Circular polarization at the level of $\sim0.6\%$ has been recently detected in the optical afterglow of GRB 121024A \cite{wiersema14}.
This is the first claim of detection of circular polarization in GRB afterglow radiation.
The same burst also shows linear polarization at a level of $\sim4\%$, implying $P^{circ}/P^{lin}\simeq0.15$. 
Assuming that synchrotron radiation is dominating the optical afterglow, both the measured $P^{circ}$ and the ratio $P^{circ}/P^{lin}$ are large as compared to expectations and they are difficult to explain: they are both expected to be not larger than $1/\gamma_e$, where $\gamma_e$ is the random Lorentz factor of the radiating electrons, and its value is estimated to be $\sim10^{4}$ at the time of polarization measurements. Anisotropies in the pitch angle distribution have been invoked as a possible explanation \cite{wiersema14}.

In this paper we estimate $P^{circ}$ in a uniform magnetic field in the plane of the shock and in a radial magnetic field, both for isotropic and anisotropic distributions of the electron velocities.
First we present the equations to derive the local circular polarization $P_0^{circ}=V_0/I_0$ (where $V_0$ and $I_0$ are the Stokes parameters)
at each point of the emitting surface.
Then, the total polarization (i.e. integrated over the emitting surface) is found. Before integration the contribution coming from each point of the surface must be weighted by the contribution $dF_\nu$ of that region to the total flux $\int{dF_\nu}$ reaching the observer.  
For a given direction of the magnetic field, the other Stokes parameters are given by $Q_0/I_0=P^{lin}_{max}\cos(2\theta_p)$ and $U_0/I_0=P^{lin}_{max}\sin(2\theta_p)$, where $\theta_p$ is the polarization position angle, $P^{lin}_{max}=(\delta+1)/(\delta+7/3)$, and $\delta$ is the slope of the electron energy distribution ($dN_e/dE\propto E^{-\delta}$). 
The local linear polarization is given by $P^{lin}_0=\sqrt{Q_0^2+U_0^2}/I_0=P^{lin}_{max}$.
To total linear polarization is given by $P^{lin}=\sqrt{Q^2+U^2}/I$, where $Q/I=\int{{Q_0\over I_0} dF_\nu}/\int{dF_\nu}$ and $U/I=\int{{U_0\over I_0} dF_\nu}/\int{dF_\nu}$. The total circular polarization will be computed as $P^{circ}=V/I=\int{{V_0\over I_0}dF_\nu}/\int{dF_\nu}$.

\section{Circular polarization}\label{sect:circ_pol}
\begin{figure}
\vskip -5 truecm
\hskip -2truecm
\includegraphics[scale=0.7]{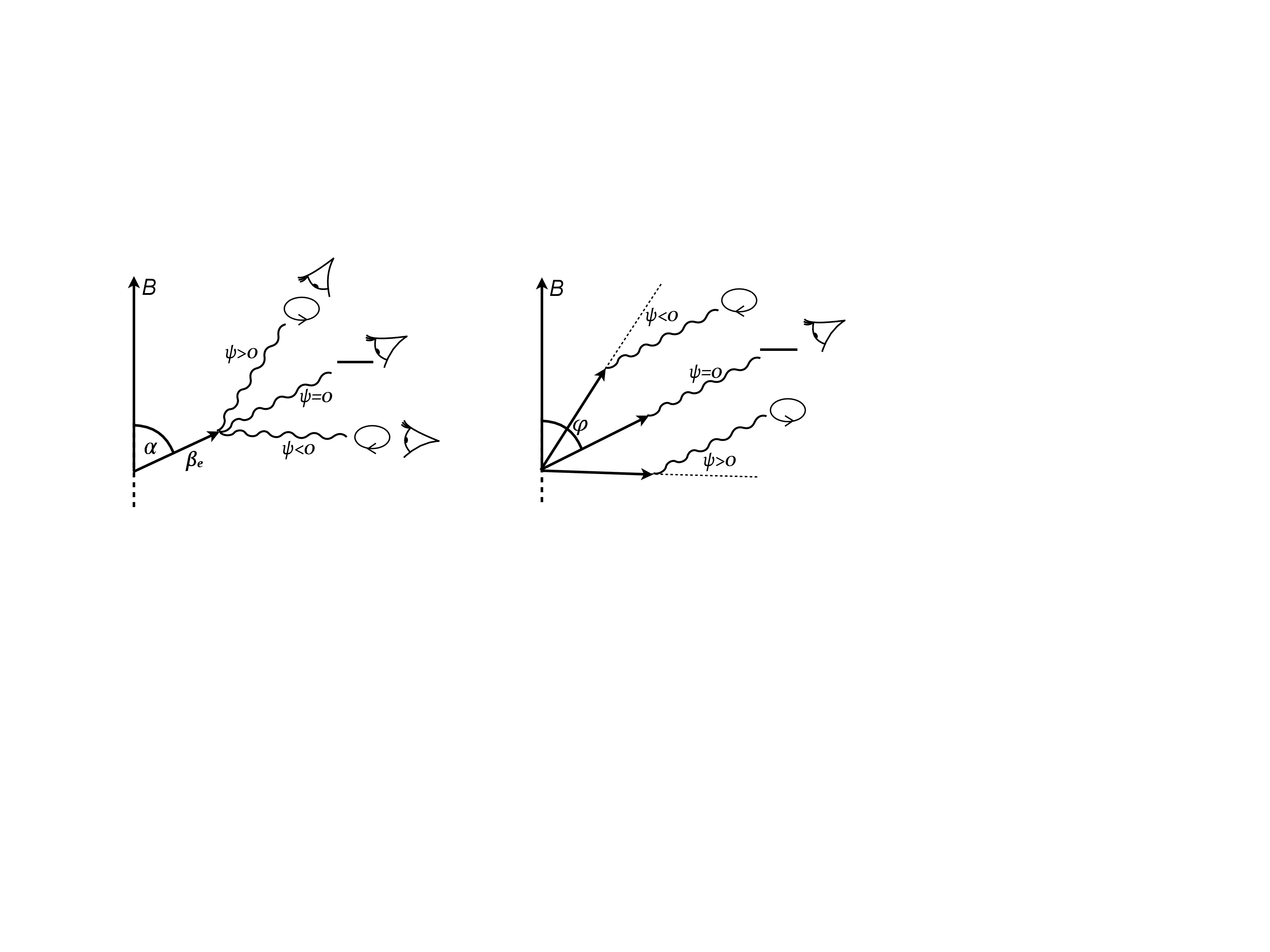}
\vskip -8.8 truecm
\caption{Left panel: polarization of the radiation emitted by an electron with pitch angle $\alpha$ as detected by different observes. The angle $\psi$ represents the angle between the electron velocity $\beta_e$ and the observer direction. Right panel: polarization detected by one observer from a distribution of electrons with different pitch angles. $\varphi$ is the angle between the magnetic field and the direction of the observer.}
\label{fig:sketch2}
\end{figure}

To understand why synchrotron radiation can be circularly polarized, we first consider emission from one single electron with a pitch angle $\alpha$ gyrating in a uniform magnetic field ${\mathbold B}$ (figure~\ref{fig:sketch2}, left panel). 
Let $\psi$ be the angle between the electron velocity ${\mathbold \beta_e}$ and the direction of the observer ${\mathbold n}$, and let consider three different observers detecting radiation from the same electron.
In general, the observed radiation is elliptically polarised, with the axes of the ellipse parallel and perpendicular to the projection of ${\mathbold B}$ on to the plane transverse to the direction of the observer ${\mathbold n}$ \cite{westfold59,legg68}. 
The major axis of the polarization ellipse is perpendicular to the projection of ${\mathbold B}$ for $|\psi|$ close to zero, but as $|\psi|$ increases the form of the ellipse changes to a circle and then to an ellipse with major axis parallel to the projection of ${\mathbold B}$. 
The direction is right-handed ($RH$) or left-handed ($LH$) according with $\psi\gtrless0$. The polarization is linear if $\psi=0$.
We note that if  $|\psi|>1/\gamma_e=\sqrt{1-\beta_e^2}$, the emission in the direction of the observer is negligible. 

Let's now consider a distribution of electrons with different pitch angles, and a single observer, located at some angle $\varphi$ with respect to the magnetic field (figure~\ref{fig:sketch2}, right panel). First we note that only electrons with pitch angle $\varphi-1/\gamma_e<\alpha<\varphi+1/\gamma_e$ contribute to the emission detected by the observer.  As follows from the previous discussion,  photons corresponding to $\psi=0$ (i.e., the one emitted by the electron with $\alpha=\varphi$) are linearly polarized. Photons for which $\psi\neq0$ instead will be elliptically polarised. The elliptical polarization is the same for photons at $\psi$ and $-\psi$, but with opposite direction of the rotation. If the number of electrons with pitch angle $\alpha=\varphi+\psi$ is different from the number of electrons with pitch angle $\alpha=\varphi-\psi$ their contributions to the total polarization do not cancel out. 
In particular, for the case of isotropic pitch angle distribution, if $\varphi<90^\circ$ the number of electrons with $\psi>0$ is larger than the number of electrons with $\psi<0$ and the total polarization is negative (as in the case in figure~\ref{fig:sketch2}, right panel). Viceversa, if $\varphi>90^\circ$ the total polarization is positive.
The local circular polarization $P^{circ}_{0}\equiv V_0/I_0$ is given by \cite{sazonov69,sazonov72}:
\begin{eqnarray}
P^{circ}_{0}=-\frac{4(\delta+1)(\delta+2)}{3\delta(\delta+7/3)}\frac{\Gamma\left(\frac{3\delta+8}{12}\right)\Gamma\left(\frac{3\delta+4}{12}\right)} {\Gamma\left(\frac{3\delta+7}{12}\right)\Gamma\left(\frac{3\delta-1}{12}\right)}\left(\cot{\varphi}+\frac{1}{\delta+2}\frac{1}{Y(\varphi)}\frac{dY(\varphi)}{d\varphi}\right)\left(\frac{3\nu_H\sin{\varphi}}{\nu^{\prime}} \right)^{1/2}
\label{eq:p0circ}
\end{eqnarray}
where $\nu_H=eB/(2\pi m_ec)$, and $\nu^\prime$ is the photon frequency. The function $Y(\alpha)$ accounts for the possible anisotropy of the velocity electron distribution: the number density per solid angle of electrons with energy $E$ and pitch angle $\alpha$ is $N(E,\alpha)\propto E^{-\delta}Y(\alpha)$.
The factor $(3\nu_H\sin{\varphi}/\nu^{\prime})^{1/2}$ is $\simeq1/\gamma_e$, where $\gamma_e$ is the random Lorentz factor of the electrons emitting at the frequency $\nu^\prime$. 

Note that in the isotropic case $dY/d\varphi=0$, and $P^{circ}_{0}$ simply reduces to:
\begin{equation}
P^{circ}_{0}= -f(\delta)P_{max}^{lin}\frac{\cot(\varphi)}{\gamma_e} 
\label{eq:pcirc_iso}
\end{equation}
\noindent
The function $f(\delta)$ assumes the value $f(\delta)\simeq2$ for $\delta$ in the range $\delta=[2.1-2.5]$.

Equation~\ref{eq:pcirc_iso} shows that locally $P^{circ}_0$ is of the order $1/\gamma_e\ll1$, i.e. it is strongly suppressed due to cancellation between electrons contributing with positive and negative polarization (as shown in figure~\ref{fig:sketch2}, right panel). 
For those configurations where the magnetic field has a given direction, $P_0^{lin}=P^{lin}_{max}\simeq0.7$, and the ratio between circular and linear polarization is also of order $1/\gamma_e\ll1$.

\subsection{Polarization maps}
\begin{figure}
\vskip -1.8 truecm
\hskip 0.8truecm
\includegraphics[scale=0.37]{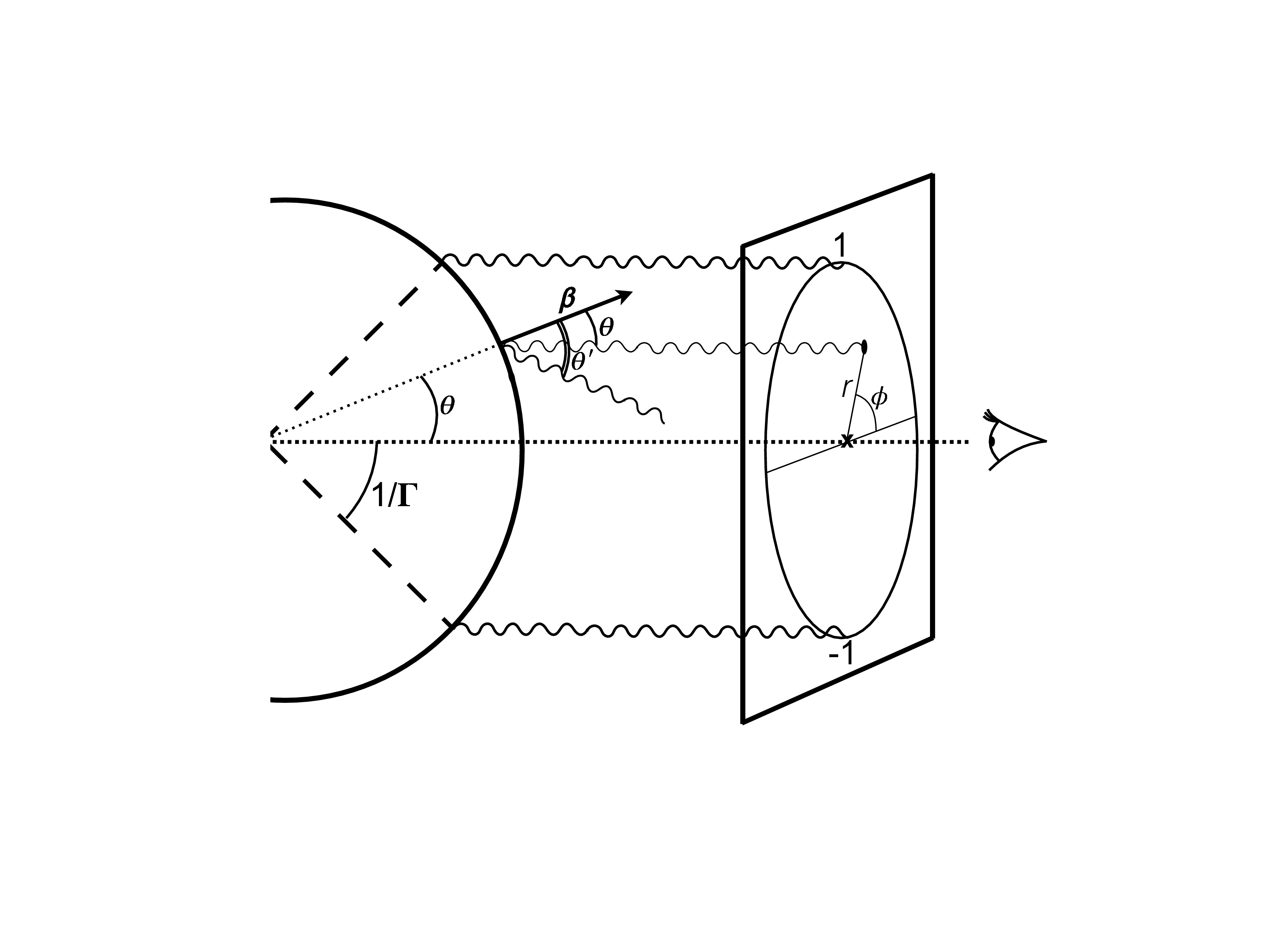}
\vskip -2.5 truecm
\caption{Representation of a polarization map: each point of the map, identified by polar coordinates $(r,\phi)$, corresponds to a pointlike region of the emitting surface, located at angular distance $\theta$ from the observer (located in the center of the map). The coordinate $r$ is equal to unity for a region located at $\theta=1/\Gamma$. The contribution from regions at $r>1$ is negligible. In the fluid frame the photon is emitted at an angle $\theta^{\prime}$ from the direction of the fluid velocity $\beta$, but in the laboratory frame it is beamed in the direction of the observer.}
\label{fig:sketch}
\end{figure}
Consider a spherical fireball radially expanding with relativistic velocity $\Gamma\gg1$ (see figure~\ref{fig:sketch}). We call $\theta$ the angle between the line of sight and the local fluid velocity: this is also the angle between the direction of the photon that reaches the observer and the electron bulk velocity $\mathbold \beta=(\Gamma^2-1)^{-1/2}$ (not to be confused with the random electron velocity $\mathbold\beta_e$ in the local fireball comoving frame). Due to beaming effects, in the fluid frame the angle between $\mathbold\beta$ and the photon is different from the one measured in the observer frame. We call this angle $\theta^\prime$ (figure~\ref{fig:sketch}).

A polarization map is the projection on the plane of the sky (perpendicular to the line of sight) of the polarization of radiation travelling in the direction of the observer, and coming from different regions of the emitting surface (figure~\ref{fig:sketch}). 
We choose to locate the observer in the center of the map.
Each point of the map is identified by polar coordinates ($r,\phi$).
The $r$-coordinate is normalised such that $r=1$ corresponds to radiation coming from a region located at $\sin{\theta}=1/\Gamma$: $r=\sin\theta\Gamma\simeq\theta\Gamma$, for $r\ll\Gamma$.
We will also use the coordinate $y$ in place of $r$, defined as $y\equiv(\theta\Gamma)^2=r^2$.
With this notation the Doppler factor is $\mathcal{D}\simeq2\Gamma/(1+y)$ (where we used the approximation $\theta\ll1$ and $\beta\simeq1$) and the angle relativistic transformations take the form $\sin\theta^\prime=2\sqrt{y}/(1+y)$.

\vskip 0.3truecm
\noindent
In the following we consider the case of both isotropic and anisotropic pitch angle distributions, and two different configurations of the magnetic field: a) uniform in the plane of the shock, and b) radial. Three different geometries of the emitting region are discussed: i) a sphere, ii) a jet with $\theta_j\Gamma=1$, and iii) a jet with $\theta_j\Gamma=1/3$.
We present the maps for the local flux-weighted polarization, and estimate the total polarization. 
Through this paper, we will perform calculations in the spectral range $\nu_m<\nu<\nu_c$ (where $\nu_m$ is the injection frequency and $\nu_c$ is the cooling frequency), 
and we use $\delta=2.5$, which implies $P^{lin}_{max}=0.72$.

\subsection{Uniform magnetic field in the plane of the shock}\label{sect:isoa_unib}
\begin{figure}
\hskip 0.1 truecm
{\includegraphics[scale=0.68]{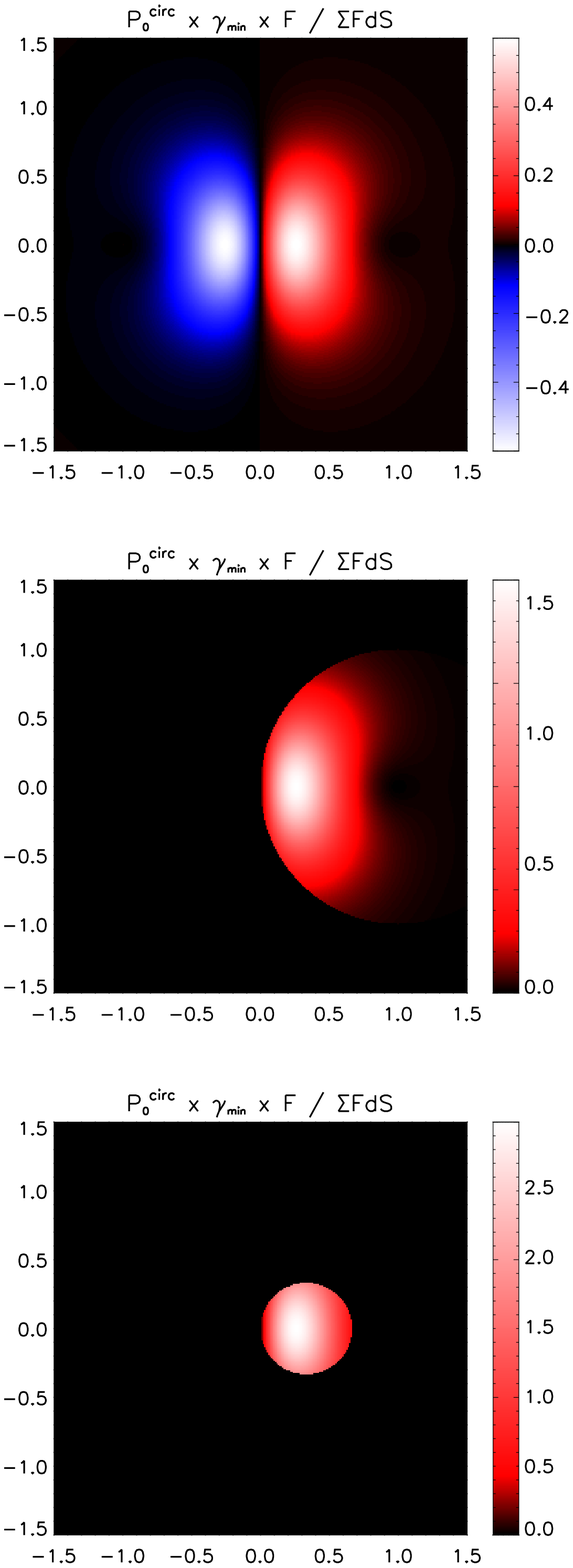}
\hskip -0.4 truecm
\includegraphics[scale=0.68]{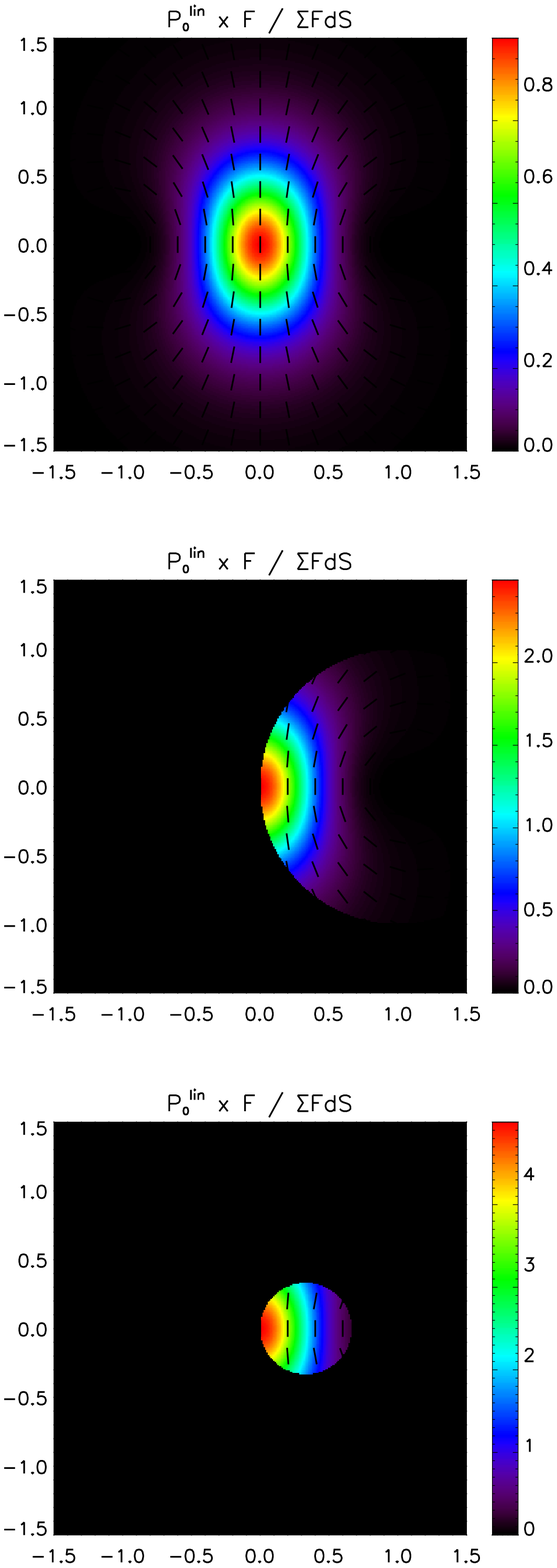}}
\caption{Maps for circular (left column) and linear (right column) polarization for the case of electrons with isotropic pitch angle distribution radiating in a uniform magnetic field in the plane of the shock. Three geometries are considered: a sphere (upper panels), a jet with $\theta_{jet}\Gamma=1$ (middle panels), and a jet with $\theta_{jet}\Gamma=1/3$ (bottom panels).}
\label{fig:isoa_unib}
\end{figure}
The equations for the angle $\varphi$ between the field and the photon (estimated in the fluid frame) and for the polarization angle can be found in \cite{granot03}.
From equation~\ref{eq:p0circ} we obtain:
\begin{equation}
P^{circ}_0\times\gamma_{e,min}=-\sqrt{\frac{\sin\varphi}{1+y}}\cot{\varphi}\,f(\delta)P_{max}^{lin}
\end{equation}
where $\gamma_{e,min}$ is the minimum Lorentz factor of those electrons radiating at the relevant frequency $\nu$.
Figure~\ref{fig:isoa_unib} shows the polarization maps for a uniform field (oriented from left to right) for the local (flux-weighted) circular (left column) and linear (right column) polarization, for the case of a sphere (upper panels), a jet with $\theta_{jet}\Gamma=1$ (middle panels) and a jet with $\theta_{jet}\Gamma=1/3$ (lower panels).
In the case of a sphere the total integrated polarization $P^{circ}=0$, and can differ from zero only for an off axis jet. 
In this case its value depends on the position of the jet axis with respect to the observer and on $\theta_{jet}$.
We derive the total polarization for each different position $(r,\phi)$ of the jet axis in the plane of the map. 
The minimum polarization is always zero and is reached when the jet axis is located at $\phi=\pm90^{\circ}$. The maximum polarization instead is around $0.35/\gamma_{e,min}$ and is reached when $\phi=0, 180^{\circ}$ and $y=0.3$ ($y=1$) for the narrower (larger) jet (these particular configurations are the ones shown in figure~\ref{fig:isoa_unib}).
The total linear polarization in the spherical case is $P^{lin}=0.61$, and ranges from 0.61 to 0.72 for the jetted geometries. 

Summarizing, in a uniform magnetic filed in the plane of the shock both the absolute value of $P^{circ}$ and the ratio $P^{circ}/P^{lin}$ can vary between 0 and $\lesssim1/\gamma_e$, and cannot explain the large values observed in the optical afterglow of GRB 121024A.

\subsection{Radial magnetic field}\label{sect:isoa_radb}
\begin{figure}
{\includegraphics[scale=0.68]{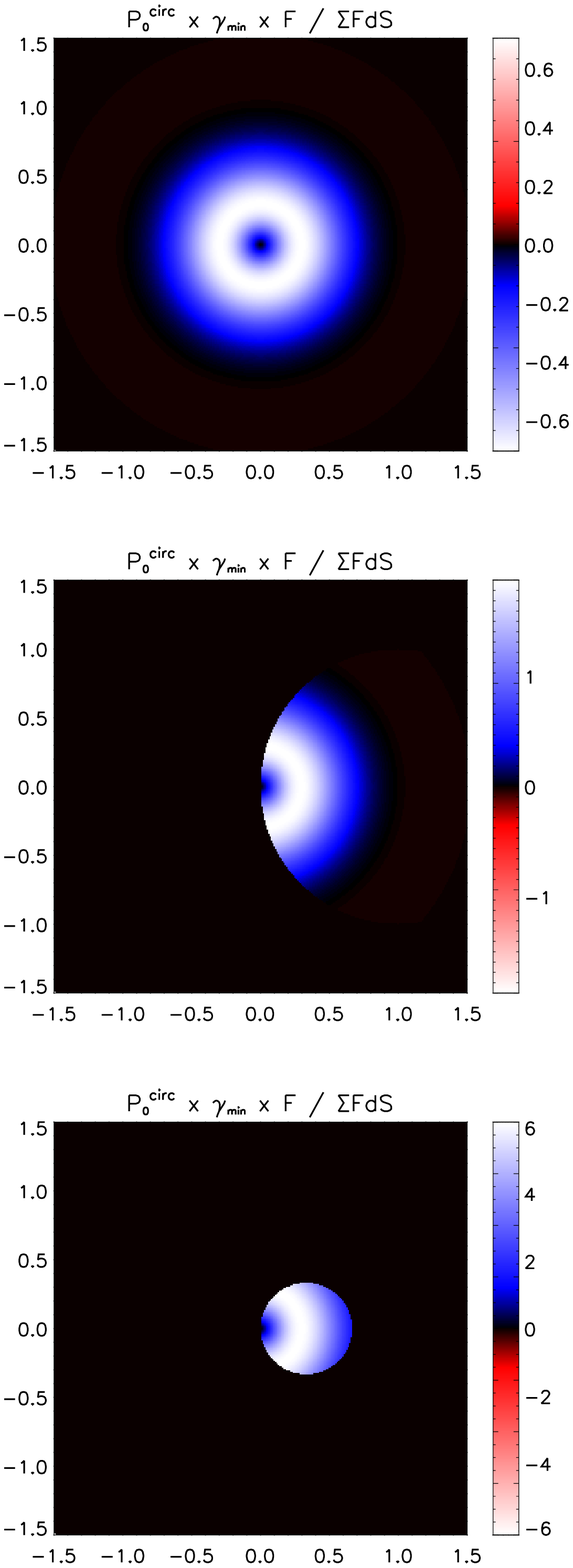}
\hskip -0.4 truecm
\includegraphics[scale=0.68]{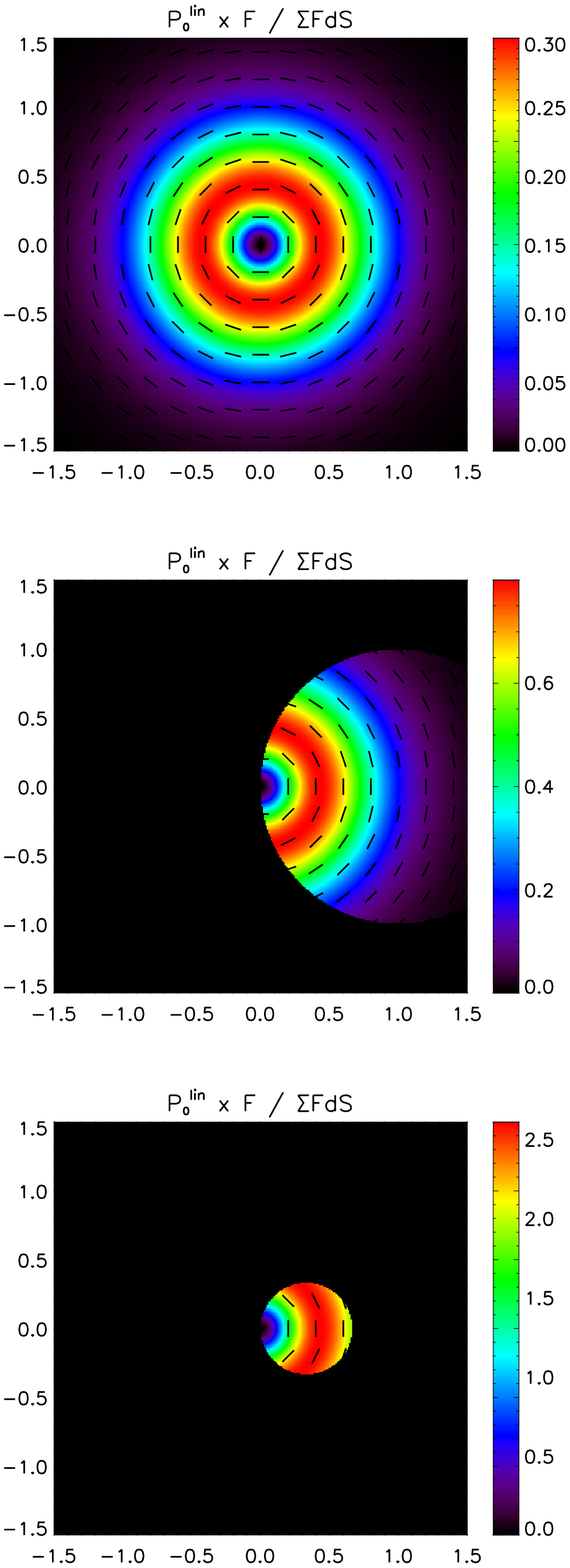}}
\caption{Maps for circular (left column) and linear (right column) polarization for the case of electrons with isotropic pitch angle distribution radiating in a radial magnetic field, perpendicular to the plane of the shock. Three geometries are considered: a sphere (upper panels), a jet with $\theta_{jet}\Gamma=1$ (middle panels), and a jet with $\theta_{jet}\Gamma=1/3$ (bottom panels).}
\label{fig:isoa_radialb}
\end{figure}
In the case of a magnetic field perpendicular to the plane of the shock, $\mathbold B$ is parallel to the fluid velocity $\mathbold \beta$, $\varphi=\theta^{\prime}$, and
$\cos\varphi=\cos\theta^\prime=\frac{1-y}{1+y}$. We obtain:
\begin{equation}
P_{0}^{circ}\times\gamma_{e,min}=-\frac{4}{3}\frac{(3y)^{1/4}}{(1+y)}\cot{\varphi}\,f(\delta)P_{max}^{lin}
\end{equation}
Polarization maps are shown in Fig.~\ref{fig:isoa_radialb}.
For $y<1$ (i.e., the region that mainly contributes to the observed flux) the angle $\varphi$ is always smaller than $90^{\circ}$ and the circular polarization is negative. 
Its contribution to the total polarization is not canceled by the positive contribution coming from the region at $y>1$, and the total polarization is $P^{circ}=-0.37/\gamma_{e,min}$ and is in the range $[0.43-1.3]/\gamma_{e,min}$ for jetted geometries.
The linear local polarization is still equal to $P^{lin}_{max}$, while the polarization angle is $\theta_p=\phi+90^\circ$  \cite{granotkonigl03,sari99b}.
When integrated over all the emission region the total polarization $P^{lin}$ is null in the spherical case, and ranges from 0 to 0.4 in the two considered jetted geometries.

This implies that for a radial $\mathbold B$, the ratio $P^{circ}/P^{lin}$ can reach very large values, and explain the observations. However, the absolute value of $P^{circ}$ is still limited to be of the order of $1/\gamma_{e,min}$.

\section{Anisotropic pitch angle distribution}
\label{sect:anisoa}
\begin{figure}
\vskip -0.35 truecm
\hskip 0.08truecm
\includegraphics[scale=0.463]{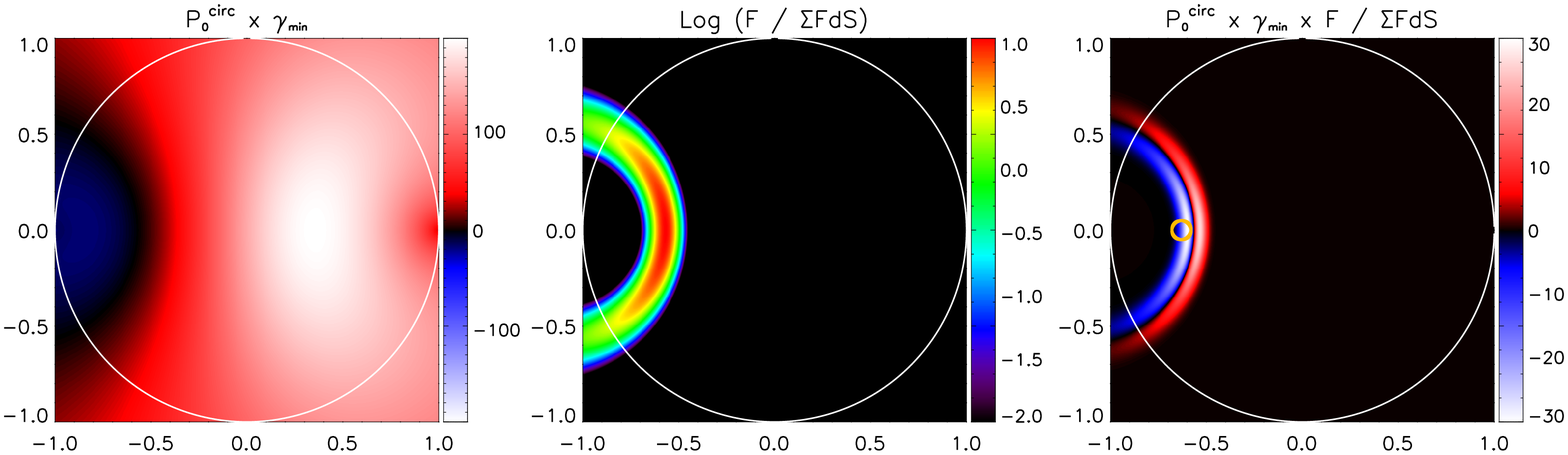}
\vskip -0.2 truecm
\caption{Maps for the local circular polarization (left), weighting factor (middle), and flux-weighted local circular polarization (right) in the case of an anisotropic gaussian distribution of pitch angles centred around $\alpha_0=30^{\circ}$ and $\sigma_\alpha=5\times10^{-2}$. The magnetic field is uniform in the plane of the shock. White circles represent the region of the sphere located at $\theta=1/\Gamma$t. The yellow circle (right panel) shows the position and extent of a jet required to obtain a circular polarisation of the order of the one observed in GRB 121024A.}
\label{fig:anisoa_unifb}
\end{figure}
We now investigate the case of an anisotropic pitch angle distribution. As an example, we consider a gaussian distribution $Y(\alpha)\propto \exp\left[-{\frac{(\alpha-\alpha_0)^2}{2\sigma_\alpha^2}}\right]$ with $\alpha_0=30^\circ$ and $\sigma_\alpha=5\times10^{-2}$, and a uniform magnetic field. A more general and accurate calculation will be presented in Nava et al. 2015 (in preparation). 
The polarisation map for this case is shown in figure~\ref{fig:anisoa_unifb} (right panel) and is the convolution of the map for the local circular polarization derived from equation~\ref{eq:p0circ} (left panel in figure ~\ref{fig:anisoa_unifb}) and the emissivity (middle panel). 
The number of electrons with $|\alpha-\alpha_0|> \sigma_\alpha$ is negligible, and the flux from these regions is suppressed due to the lack of electrons with such pitch angles. Only electrons with $|\alpha-\alpha_0|\lesssim\sigma_\alpha$ contribute to the emission.
This is why the radiation can reach the observer only from a small annular region for which the corresponding pitch angle is around $\alpha_0$ (see figure~\ref{fig:anisoa_unifb}, right panel).
For $\alpha>\alpha_0$ ($\alpha<\alpha_0$) the local polarization is positive (negative).  
Using $|\alpha-\alpha_0|\simeq\sigma_\alpha$ we derive $P_{0}^{circ}\gamma_{e,min}\simeq\sigma_\alpha^{-1}$, that gives local values close to the measured one.
However, when integration over the emitting surface is performed, partial cancellation from contributions with opposite rotation directions takes place. 
We find that an extra factor $\sigma_\alpha$ arises from the integration, and the total circular polarization is again of the order of $1/\gamma_{e,min}$. 
This can be avoided only by considering a jet with an unrealistically small opening angle ($\theta_{jet}\sim1/20\Gamma$), 
whose position allows the observer to see only the part of the emitting region contributing with positive (or negative) polarization (see figure~\ref{fig:anisoa_unifb}, right panel, yellow circle).

We conclude that the circular polarization observed in GRB 121024A cannot be explained as intrinsic to the radiation, if this is dominated by synchrotron from external shocks.

\acknowledgments
This work was supported by a Marie Curie IEF (LN), by ERC grants GRBs (TP) and GRB/SN (EN), by a grant from the Israel ISF - China NSF collaboration (TP), by a grant from the Israel Space Agency (TP and EN), and by the I-Core Center of Excellence in Astrophysics.

\end{document}